\documentclass[twocolumn,floatfix,prl,showpacs,amsmath]{revtex4}

\usepackage{graphicx}
\usepackage{graphics}
\usepackage{epsfig}
\usepackage{epstopdf}
\usepackage{bm}
\usepackage[usenames]{color}

\begin{document}

\preprint{LA-UR-04-5566}

\title{Energy dependence on fractional charge for \\ strongly interacting subsystems}

\author{Steven~M.~Valone}
\affiliation{Materials Science and Technology Division, Los Alamos
National Laboratory,
Los Alamos, New Mexico 87545 and \\
Department of Physics and Astronomy, University of New Mexico,
Albuquerque, New Mexico 87131}

\author{Susan~R.~Atlas}
\affiliation{Center for Advanced Studies and
Department of Physics and Astronomy, \\ University of New Mexico,
Albuquerque, New Mexico 87131
\vspace*{0.05in}}

\date{\today}

\begin{abstract}
The energies of a pair of strongly-interacting subsystems with
arbitrary noninteger charges are examined from closed and open
system perspectives.  An ensemble representation of the charge
dependence is derived, valid at all interaction strengths.
Transforming from resonance-state ionicity to ensemble charge
dependence imposes physical constraints on the occupation numbers in
the strong-interaction limit.  For open systems, the chemical
potential is evaluated using microscopic and thermodynamic models,
leading to a novel correlation between ground-state charge and an
electronic temperature.
\end{abstract}

\pacs{71.15.-m, 31.10.+z, 34.20.-b, 34.70.+e}
\maketitle

Associating fractional charges with individual atoms is a habitual
part of our every-day thinking about condensed and molecular matter.
Indeed, characterizing the energetics of such systems in terms of
dynamically-evolving charges is now recognized as key to
understanding the atomic-scale behavior of complex processes ranging
from alloying to motor protein function to molecular logic gate
operation \cite{CTRef}. Compounding the difficulty of the problem is
that the instantaneous redistribution of charge occurs under the
influence of strong interactions among subsystems of atoms and
molecules.

Historically, it was assumed that quadratic electrostatic
interactions constitute a reasonable representation of the energy
dependence on fractional charge \cite{QEq}, {\it independent} of the
strength of interaction.  However, the inadequacy of
the historical assumption had been highlighted by the work of Perdew {\it
et al.}~(hereafter referred to as PPLB) \cite{PPLB}, where a linear dependence was found for weakly interacting subsystems.  This was found by considering an {\it open}
subsystem that was allowed to weakly interact and exchange electrons with
a reservoir of electrons \cite{PerdewNATO}. For a diatomic molecule,
the weak interaction restriction implies that the theory is valid
only at large internuclear separations $R$.  PPLB considered an atom A as a subsystem
in its neutral state with $M$ electrons and energy $E_M$ that
becomes anionic by fractional charge $q$, where the anionic state
with $M+1$ electrons has energy $E_{M+1}$. The energy $E_{\rm A}(q)$
was shown to be the ensemble average
\begin{eqnarray}
    E_{\rm A}(q) = E_M + \omega \, (E_{M+1} - E_M) \, ,
    \label{eq:PPLBe}
\end{eqnarray}
with $\omega \sim \omega_{\rm PPLB} = -q \ge 0$. This result
established a seminal extension of density functional theory (DFT)
to fractional numbers of electrons \cite{Persp}.  The energy $E_{\rm
A}(q)$ is manifestly {\it linear} in $q$, in contrast to the
historically-assumed quadratic dependence.  A direct consequence of
Eq.~(\ref{eq:PPLBe}) is that the associated microscopic chemical potential $\mu = -dE_{\rm A}(q)/dq$
exhibits a discontinuity with respect to charge $q$
\cite{PPLB}.  The veracity of both limits has been
confirmed numerically by
Cios{\l}owski and Stefanov \cite{CioStef}.  In this Letter, we construct a novel, {\it analytic} model
explicitly linking the weak (linear) and strong (quadratic)
interaction limits, and characterize the corresponding behavior of
$\mu$ with respect to $q$.  As discussed below, this requires the
definition of interacting ``atom-in-molecule" (AIM) subsystems and
associated fractional charges.   While these definitions are not unique
\cite{PAN05}, and many physically-reasonable alternatives are
possible \cite{HarrisCharge, GhosezPerov, CohenWass, WuVoorhis}, our
results are independent of the details of a specific AIM approach.

Previous attempts to extend PPLB's formal results to moderate and
strong interactions have recast the problem in terms of charge
resonances \cite{CT-EVB, Nal-KS, ToddM, ToddMII} of a {\it closed}
system composed of A$+\cal{R}$, where $\cal{R}$ represents the
electron reservoir. In the charge resonance view, one supposes that
special wavefunctions can be constructed such that the charges on
the subsystems are integers \cite{Warshel,MullikenDi,
PhillipsRMP,CoulsonRedei, AdelHersch}. Interpreting PPLB in the
language of resonances, the state designated as $M$ for A
corresponds to both A and $\cal{R}$ being
neutral (covalent resonance), while $M+1$
corresponds to A being anionic and $\cal{R}$ being
cationic (ionic resonance), leaving the total system neutral.

Extending this picture to the {\it intermediate interaction} regime
\cite{CT-EVB, fn1}, ${\cal R}$ ceases to be a structureless
reservoir of electrons, but must instead be viewed as a structured
subsystem B at a finite separation $R$ from A.
This subsystem may be an atom, molecule, or bulk material. Although
several two-state valence bond models have been developed in this
limit \cite{CT-EVB,Nal-KS,ToddM,ToddMII}, previous attempts to
extend the valence bond approach to the strong interaction limit
have proven unsuccessful.

\begin{figure}
\begin{center}
     \includegraphics[scale=.2]{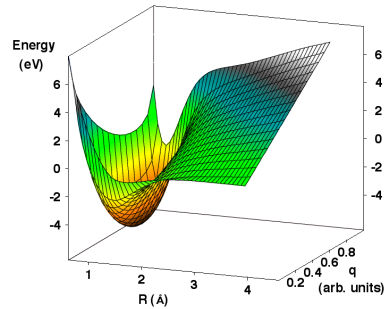}
     \caption{(Color) Energy of a heteronuclear diatomic molecule as a
     function of bond length and charge as modeled by
     Eq.~(\ref{eq:EnsRepE}).  The charge dependence is linear as the molecule dissociates \cite{PPLB}, while being quadratic near the ground-state charge when $R$ is near
equilibrium values \cite{QEq, CioStef}.  The constituent atoms interact strongly near equilibrium.}
     \label{fig:E_HF_surf}
\end{center}
\end{figure}

\begin{figure}
\begin{center}
    \includegraphics[scale=0.22]{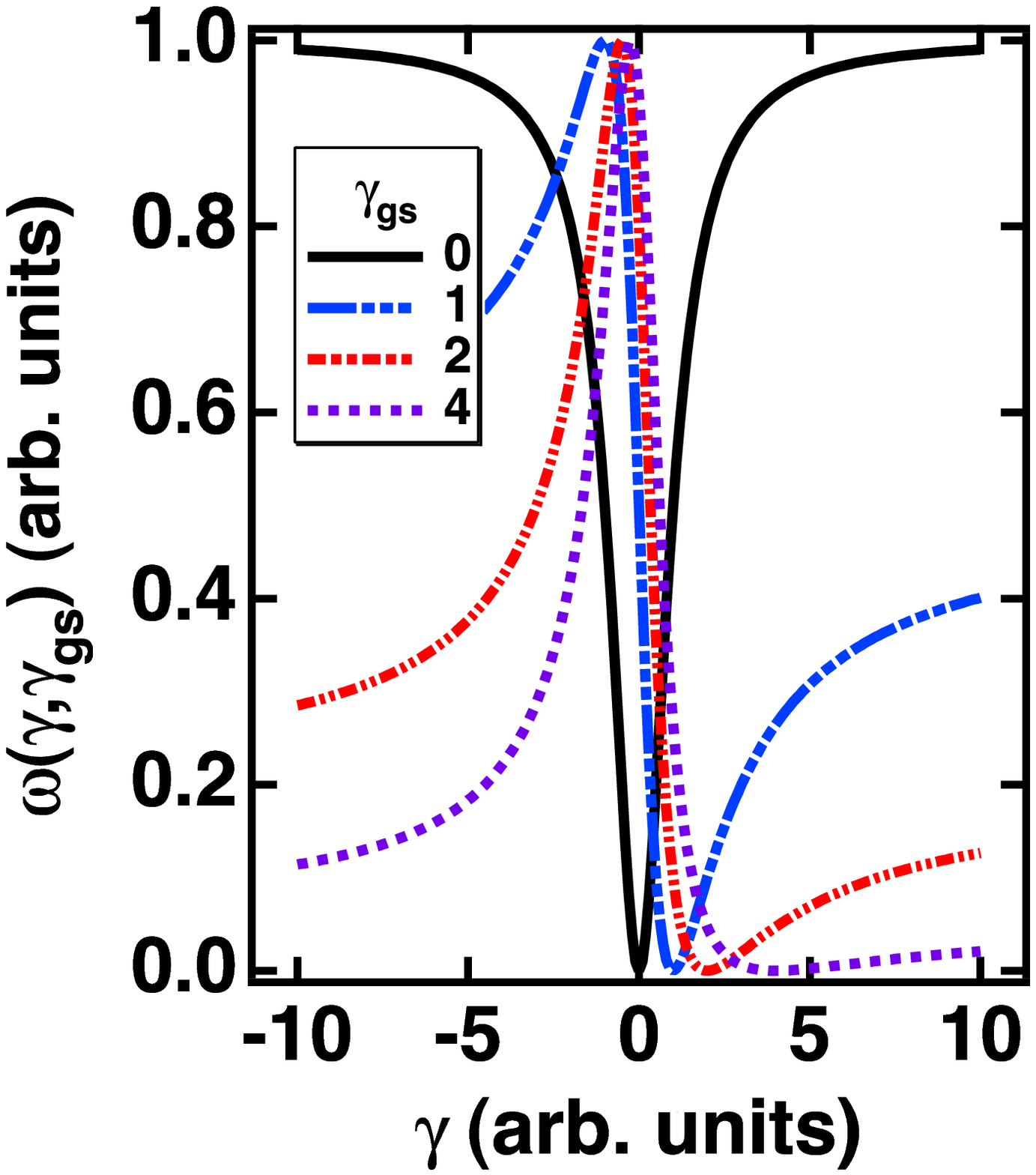} (a)
    \includegraphics[scale=0.20]{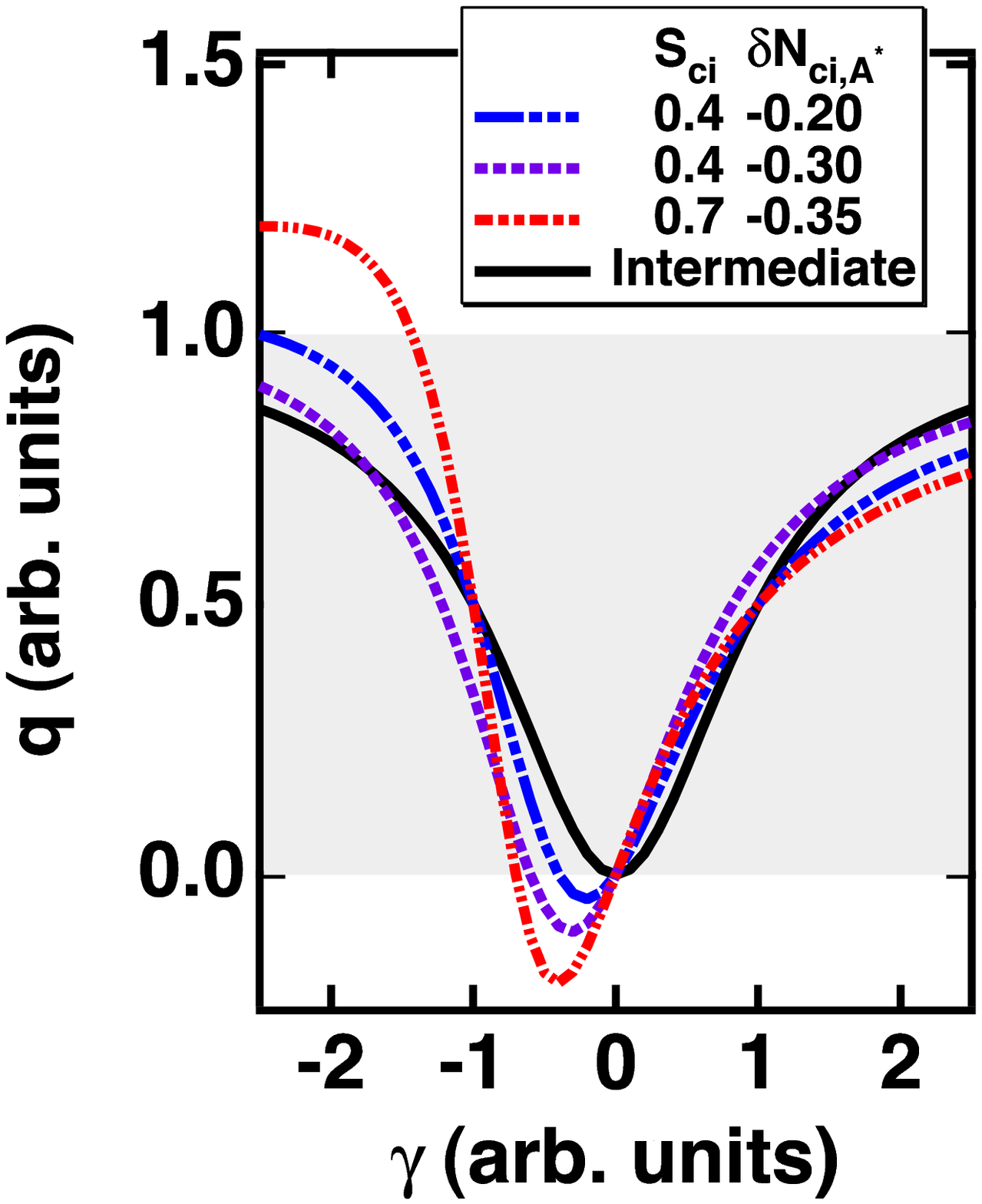} (b)
    \caption{(a) (Color online) Intermediate limit of
    Eq.~(\ref{eq:EXACTocc}).  The extrema correspond to $\gamma = \gamma_{\rm
    gs}$ and $\gamma = \gamma_{\rm xs} = -1/\gamma_{\rm gs}$.
    (b) Relationship between charge and ionicity in the strong interaction regime
    for the two-state model and the atom-in-molecule model of Ref.~\cite{CT-EVB}.  $S_{\rm ci}$ and $\delta N_{\rm ci}^{*}$
    are AIM parameters entering the $\gamma-q$ relation in the strong interaction regime.  The gray
    zone corresponds to the physical range of $q$ between 0 and 1. }
    \label{fig:omega-NDOL}
\end{center}
\end{figure}

A complete generalization of PPLB to the strong
interaction limit can be attained by transforming from the resonance-state
basis to an ensemble or spectral representation \cite{PerdewNATO}. Let $\hat{H}$
denote the hamiltonian  for the closed system AB with eigenstates
$\{\Psi_k\}$ and eigenenergies $\{E_k\}$.  Let $\Psi$ be any arbitrary trial wavefunction with
$\langle \Psi | \Psi \rangle  = 1$.  The variational energy can then be expressed in a spectral representation,
\begin{equation}
    \langle \Psi | \hat{H} | \Psi \rangle = \sum_k \omega_k E_k \, ,
    \label{eq:EnsVarPrin}
\end{equation}
where the $\omega_k = | \langle \Psi | \Psi_k \rangle |^2$ are
occupation numbers.  Now we express $\Psi$ as a linear combination of resonance states $\{
\psi_i\}$ with coefficients $\{c_i\}$, such that $\Psi = \sum_{i} c_{i} \psi_{i}$. The eigenvectors $\bm{C}_{k}$ are the specific values of coefficients corresponding to the eigenstates. The basis rotation from the $\{
\psi_i\}$ to the $\{\Psi_k\}$ in terms of the $\bm{C}_{k}$ lead to
occupation numbers
\begin{equation}
    \omega_k = |\sum_{i}  c_{i}^{*} {C}_{ik} |^2 \, .
    \label{eq:EnsVarPrinOcc}
\end{equation}
These occupation numbers give rise to piecewise linearity in the energy when AB dissociates \cite{PPLB,YZA-Frac}. Specializing to a two-state model, we define the {\it ionicity}
$\gamma$ as the ratio $c_{1}/c_{0}$, where $c_{0}$ and $c_{1}$ are the
coefficients for the covalent and ionic resonances, respectively. Let $\gamma_{\rm gs}$ and $\gamma_{\rm xs}$
denote the eigenionicities corresponding to the ground- and excited-state values of $c_{1}/c_{0}$, at some chosen separation $R$ \cite{CT-EVB}. The dependence of the energy $E(\gamma)$ on an externally-imposed ionicity $\gamma$ is then given by
\begin{eqnarray}
    E(\gamma; R) &=& E_{\rm gs}(R)  \nonumber \\
    && \hspace{-40pt} + \, \omega(\gamma; \gamma_{gs}(R), \gamma_{xs}(R))
    \left(E_{\rm xs}(R) - E_{\rm gs}(R) \right) \, ,
    \label{eq:EnsRepE}
\end{eqnarray}
with
\begin{eqnarray}
    \omega(\gamma; \, \gamma_{\rm gs},\gamma_{\rm xs}) =
    \frac{(\gamma - \gamma_{\rm gs})^2}{(\gamma - \gamma_{\rm
    gs})^2 + \big( \frac{1 - \gamma_{\rm gs}^2}{\gamma_{\rm xs}^2 -
    1} \big) \left( \gamma - \gamma_{\rm xs} \right)^2 } \, .
    \label{eq:EXACTocc}
\end{eqnarray}
This occupation number governs how the energy of AB changes when
its ionicity is forced to deviate from its ground-state value (Fig.~\ref{fig:E_HF_surf}).
Importantly, this result is independent of the strength of the
interaction.  As is necessary physically, $\omega = 0$ when $\gamma
= \gamma_{\rm gs}$, and $\omega = 1$ when $\gamma = \gamma_{\rm
xs}$.  Moreover, both extrema are represented, as the derivative of
$\omega$ with respect to $\gamma$ is zero when evaluated at either
$\gamma_{\rm gs}$ or $\gamma_{\rm xs}$ (Fig.~\ref{fig:omega-NDOL}a).

\begin{figure}[b]
\begin{center}
    \includegraphics[scale=0.18]{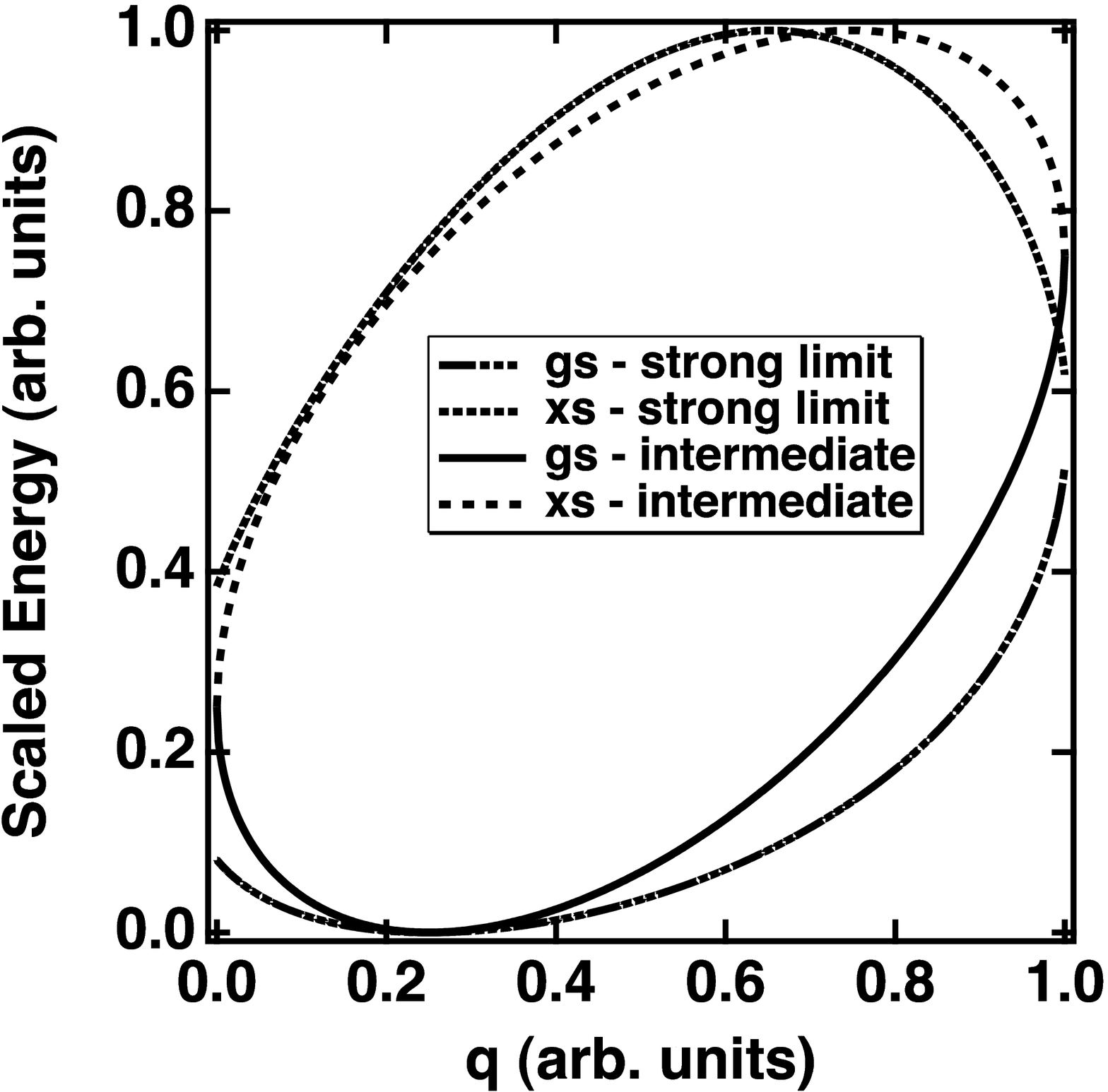}(a)
    \includegraphics[scale=0.18]{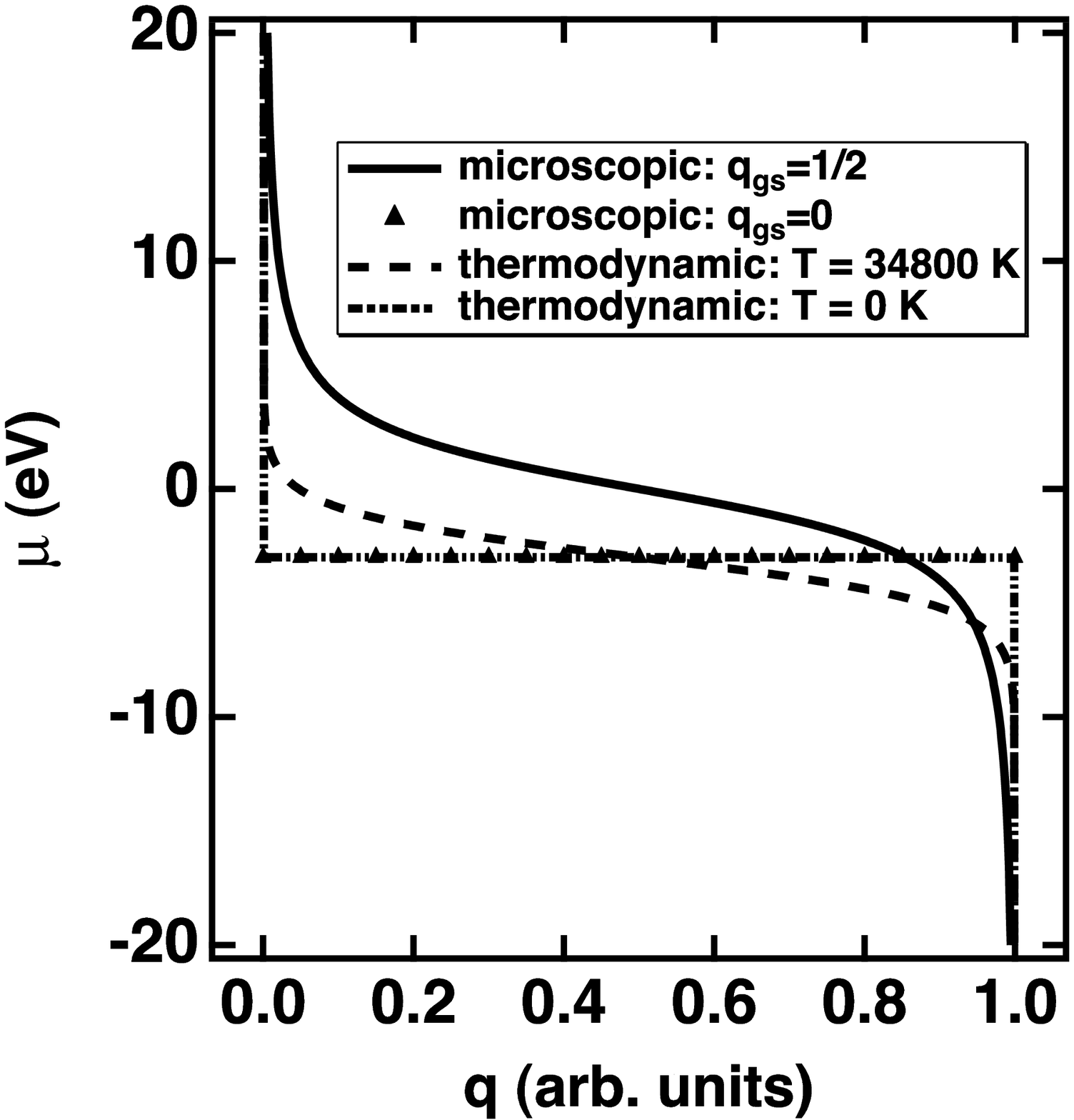}(b)
    \caption{(a) $\omega(q)$, corresponding to branches of the energy surfaces, Eq.~(\ref{eq:EnsRepE}), in the strong and intermediate limits for each root of the $\gamma-q$ relationship: $S_{\rm ci} = 0.25$ and $\delta N_{\rm ci}^{*} = -0.20$ (strong); $S_{\rm ci}=0$ and $\delta N_{\rm ci}^{*}=0$ (intermediate). $q_{\rm gs} = 0.25$ in both cases.  Gaps form at integer charges.  (b) Two-state microscopic and thermodynamic models of the chemical potential.
    The microscopic model is plotted for $q_{\rm gs} = 0$ and $1/2$, and
    the thermodynamic model for $T = 0$ and 34800 K.
    }
    \label{fig:CP-models}
\end{center}
\end{figure}

Equation~(\ref{eq:EXACTocc}) is expressed in terms of the ionicity
$\gamma$.  In order to complete the model it is necessary to
eliminate this quantum mechanical parameter in favor of the
physical charge $q$ \cite{Fano}.  This can be accomplished through
an AIM decomposition \cite{CT-EVB, CioStef}.  The
relation between $\gamma$ and $q$ in the strong interaction regime
is illustrated in Fig.~\ref{fig:omega-NDOL}b. It reduces to
$\gamma = \pm \sqrt{q/(1-q)}$ in the intermediate interaction regime
\cite{CT-EVB, CoulsonRedei, PhillipsRMP}, yielding $\omega(q)\,\sim\,\omega_{\rm int}(q, q_{\rm gs}(R))$ = $( \sqrt{(1-q_{\rm gs}(R))\, q} \pm \sqrt{q_{\rm gs}(R) \, (1-q)} \,
)^2$ \cite{CT-EVB}. Although details of the $\gamma-q$
relation will be affected by the specific choice of AIM charge
definition, we can immediately identify two quite
general consequences of the transformation to $q$. First, in
specifying the $\gamma-q$ relation, choosing a root, and restricting the charge to the physical range between 0 and 1, some values of the ionicity are excluded.  This outcome was anticipated by Pan {\it et al.~}\cite{SahniPM}.  As a result,
$\omega(q)$---in contrast to $\omega(\gamma)$---may no longer be a
proper occupation number, since it does not span the full range
from 0 to 1.  Second, the presence of the two roots results in two
branches in the $E(q)$ surface (Fig.~\ref{fig:CP-models}a; only one was shown in
Fig.~\ref{fig:E_HF_surf}), provided that the gap in Eq.~(\ref{eq:EnsRepE}) is nonzero.  The branches are degenerate for all $q$ in the weak limit, degenerate only at the integer charges in the intermediate limit, and completely nondegenerate in the strong limit.

As the driving force for such charge transfers, the concept of chemical potential  presupposes an ability to specify subsystems.  However, this concept requires the identification of the subsystem energies, as well as charges.  As with the AIM charge, the subsystem energy is not uniquely defined.  Regardless, certain general features are valid for {\it any} definition of  subsystem energy, provided that the total energy of the parent closed system is preserved.  That is, for closed system energy $E_{\rm AB}$, the open subsystem energies $E^{*}_{\rm A}$ and $E^{*}_{\rm B}$ must sum to $E_{\rm AB}$:  $E^{*}_{\rm A} + E^{*}_{\rm B} = E_{\rm AB}$.  (Asterisks indicate open-system status with respect to energy and electron transfer.)  This proviso is necessary because we require the stationary properties of the eigenstates of the closed system.

Now, if $E_{\rm AB}$ is either the ground or an excited state energy, it must follow that the subsystem energy definition applies to these as well, thereby defining $E_{\rm A,gs}^*$ and $E_{\rm A,xs}^*$.  Thus, for subsystem ${\rm A^*}$, its energy $E_{\rm A}^*$ with an arbitrary charge $q$ becomes
\begin{eqnarray}
    E_{\rm A}^*(q) &=& E_{\rm A,gs}^* + \omega(q) \, ( E_{\rm A,xs}^* - E_{\rm A,gs}^*) \, ,
    \label{CP.0}
\end{eqnarray}
where $\omega(q)$ comes from Eq.~(\ref{eq:EXACTocc}) and the exact $\gamma-q$ relation.  An analogous expression holds for ${\rm B^*}$.  It is transparent that the sum of the two AIM energies recovers the total energy of the closed system, independent of the details of the chosen subsystem energy definition.  Likewise, if $E_{\rm AB}$ represents a small change from a ground-state energy, then one concludes that $-dE_{\rm A}^*/dq = dE_{\rm B}^*/dq \equiv \mu^*$, where $\mu^*$ is the microscopic chemical potential.  Differentiating Eq.~(\ref{CP.0}), and using the simpler intermediate regime with the negative root of the $\gamma-q$ relation, yields
\begin{eqnarray}
    \mu^*(q) &=& -( E_{\rm xs}^* - E_{\rm gs}^*) \, \nonumber \\
    &\times& (1 - 2 \, q_{\rm gs} \pm  (1 - 2 \, q) \sqrt{\frac{(1-q_{\rm gs}) \,
    q_{\rm gs}}{(1-q) \, q }} ) \, ,
    \label{CP.3}
\end{eqnarray}
where the $R$ dependence and atom subscript have been suppressed for clarity (Fig.~\ref{fig:CP-models}b).  $\mu^*$ is not defined at integer charges, thereby preventing the subsystem from transferring a full charge.  This behavior comes from the degeneracy in the branches of $E(q)$ at integer $q$ and is intimately related to the derivative discontinuity found by PPLB.  At $q = q_{\rm gs}$ with the negative root, $\mu^*=0$, in accordance with our zero of energy.  Nonetheless, chemical potential equalization \cite{SAN51} does hold.

By contrast, for strong interactions with the exact $\gamma-q$ relationship, $\mu^*$ is continuous at 0 and 1.  Smoothing of the chemical potential arises directly from the movement of the derivative discontinuities to noninteger q outside of the physical range of [0, 1] (Fig.~\ref{fig:CP-models}a).  The subsystem can now transfer a full charge, but the driving force may be very large."

To complete the discussion of chemical potential and in analogy to PPLB,  we compare microscopic (Eq.~(\ref{CP.3})) and thermodynamic definitions \cite{ChatCedParr, ToddM, APN02}.  The thermodynamic definition is relevant because the number of states in the spectral representation, Eq.~(\ref{eq:EnsVarPrin}), becomes exponentially large with system size.  The grand canonical ensemble is most appropriate here \cite{PPLB, GYH68}.  For an ensemble characterized by chemical potential $\mu$ and inverse temperature $\beta = 1/k_{\rm B}T$, the grand partition function is $\sum_k \exp(\beta \mu N_{k}^* - \beta E_{k}^*)$, where $N_{k}^*$ denotes the number of electrons in state $k$ with energy $E_{k}^*$ \cite{Tolm}.  Considering again a two-state model, the average electron number is $\langle N^* \rangle =  N_{\rm gs}^* + (N_{\rm xs}^* - N_{\rm gs}^*) \exp(\beta \mu (N_{\rm xs}^* - N_{\rm gs}^*) - \beta (E_{\rm xs}^* - E_{\rm gs}^*))/(1+\exp(\beta \mu (N_{\rm xs}^* - N_{\rm gs}^*) - \beta (E_{\rm xs}^* - E_{\rm gs}^*)))$, whose solution for $\mu = \mu(q)$ with $q = N_{\rm gs}^* - \langle N^{*} \rangle$ \cite{PPLB, PerdewNATO, GYH68} is
\begin{eqnarray}
    \mu(q) = \frac{E_{\rm xs}^* - E_{\rm gs}^* + k_{\rm B} T \ln (\frac{q}{N_{\rm gs}^* - N_{\rm xs}^* - q} )}{N_{\rm xs}^* - N_{\rm gs}^*} \, .
    \label{CP.4}
\end{eqnarray}
At $T = 0$ K, $\mu = (E_{\rm xs}^* - E_{\rm gs}^*)/(N_{\rm xs}^* - N_{\rm gs}^*)$ for all $q$.  By comparing Eqs.~(\ref{CP.3}) and (\ref{CP.4}), we observe that the $T = 0$ K result corresponds to $q_{\rm gs} = 0$ (Fig.~\ref{fig:CP-models}b).  On the other hand, a temperature of 34800 K corresponds to $q_{\rm gs} = 1/2$,
if energy $E_{\rm xs}^* - E_{\rm gs}^* = 3$ eV and $N_{\rm xs}^* - N_{\rm gs}^* = -1$ are chosen.

Because there are no other degrees of freedom with which to
equilibrate, it is clear that $T$ corresponds to an
electronic temperature.  In fact, this is the grand canonical (open
system) analog of the electronic temperature introduced by Kohn
\cite{KOH86} in the context of calculating excitation energies using
closed system ensemble DFT \cite{OGK88, GOKII}.  The corresponding
\textit{canonical} ensemble is defined in terms of a spectral
representation, as here, and is characterized by a temperature
$\theta$.  $\theta$ is defined implicitly through a self-consistent
relation equating the total entropy of the system to the integrated
local entropy of the system's electron density distribution, in turn
defined through a set of state- and temperature-dependent Kohn-Sham
equations \cite{KOH86,OGK88}.  There is thus a direct analogy
between our electronic temperature for an open system with charge
transfer, and that of Kohn {\it et al.}, even though they restricted
their attention to polarization excitations of a closed system with
fixed total charge.  In both treatments, the temperature corresponds
to deviation of the charge distribution away from the ground state
distribution.  From a thermodynamic point of view, the excitations
lead to an increase in entropy.  The correspondence observed between
$T$ and $q_{\rm gs}$ stems from the fact that both quantities encode
information about the energetics of the microscopic states of the
interacting subsystems.  Equivalently, both $T$ and $q_{\rm gs}$
reflect the strength of the coupling between a subsystem and a reservoir.

In conclusion, we have shown how charge dependencies in a pair of
subsystems can be described by appealing to an ensemble
representation of the closed system energy, even in the strong
interaction limit.  The analytical form of the charge-dependent
system energy is determined through a basis rotation involving the
resonance states that embody charge transfer between subsystems.
When a decomposition method is applied to closed system
eigenenergies, charge-dependent, open system energies can be
defined, leading to a microscopic model of the chemical potential.
When the subsystems interact strongly, the derivative
discontinuities in the energy found by PPLB move to noninteger
charges that are outside of our physically allowed range, leading to
continuous chemical potentials at the integer charges.  When
compared with a thermodynamic model of the chemical potential, a
correlation is observed between the ground-state charge and an
electronic temperature, thus defining complementary measures of the
interaction strength.

The ensemble variational energy defined here may be regarded as the
wavefunction predecessor of the excited-state density functionals of
Gross {\it et al.}~\cite{GOKII}.  However, the formal connection of
our strongly-interacting, open system results to density functional
theory remains an open question, requiring the construction of a
variational principle for open system excited states.  To the best
of our knowledge such a variational principle has not yet been
established.


The work of SMV was performed in part at Los Alamos National Laboratory under the auspices of the US~Department of Energy, under contract No.~W-7405-ENG-36, and funded through the Advanced Fuel Cycle Initiative.  SMV thanks the University of New Mexico, Department of Physics and Astronomy for its hospitality during the 2003-2004 academic year.  The authors thank the National Science Foundation for support of this work under grants DMR-9520371 (SRA) and CHE-0304710.  SMV dedicates this work to his sister Susan.


\begin{thebibliography}{99}
\bibitem{fn1} ``Strength'' is defined through the energies of
and overlaps between the charge resonances as defined in
Ref.~\onlinecite{CT-EVB}:  ``Intermediate'' means zero overlap
but nonzero resonance energy.  ``Weak'' means that
both quantities are zero.
\end{thebibliography}
\end{document}